\documentclass [dvips, 12pt] {article} 
\usepackage {latexsym, amssymb, amsmath, graphicx}
\usepackage {pifont}
\newcommand*{\Approx}[1]{\ensuremath {\approx #1}}

\newcommand {\cd} {\normalsize \textcircled {\em \scriptsize d}}
\newcommand {\cu} {\normalsize \textcircled {\em \footnotesize u}}
\newcommand {\cs} {\normalsize \textcircled {\em \footnotesize s}}

\newcommand {\cdb} {\normalsize \textcircled {\em \scriptsize   $\bar{d}$}}
\newcommand {\cub} {\normalsize \textcircled {\em \footnotesize $\bar{u}$}}
\newcommand {\csb} {\normalsize \textcircled {\em \footnotesize $\bar{s}$}}

\begin {document}

\begin {center}
{\Large The exact interpretation of mass differences 
	  in upper part of the charmonium spectrum}
     \par\bigskip 
{\large Oleg~A.~Teplov}\par\smallskip 
Institute of metallurgy and materials science of the Russian Academy of Science, 
Moscow.
\par e-mail:\quad oleg.a.teplov@gmail.com  \quad or \quad  teplov@ultra.imet.ac.ru 
\end {center}

\begin{abstract}

Mass differences of the charmonium mesons $\psi$(3770), $X$(3872),
$\chi$(3929),  $X$(3940), $\psi$(4040), $\psi$(4160), $X$(4259) 
and $\psi$(4415) have been studied.
The analysis of data has shown, 5 of 7 mass differences 
with $\psi$(4415) are precisely interpreted by using of harmonic quarks 
and their complete oscillators.
Upper part of the charmonium spectrum consists of two groups. 
Except for two particles ($\psi$(3770)
and $\psi$(4040)) connected among themselves, the others 5 particles 
are connected with a meson $\psi$(4415). 
The levels are separated by elementary groups of harmonic quarks
and comleted oscillators. Experimental and calculated data are in the good consent. 
The masses of mesons $\psi$(4415), $X$(4260), $\psi$(4160), $X$(3943) 
and $\chi$(3929) were recalculated. The second group demonstrates that harmonic quarks 
can also well work below threshold $D\bar{D}$.

\end{abstract}

\section {Introduction}
\quad In~\cite{my1} 
was shown, that the mass differences
of the $D$ mesons are rigorously quantized by the rest energies
of harmonic quarks. So the mass difference of mesons $D(2007)^0$ and $D^0$
 corresponds to neutral quark group \cd$u^0$\cdb\, with the rest energy 142.124 MeV.\\
The $D_s^{*\pm}$ - $D_s^{\pm}$ mass difference is quantized 
by the simplest quark reaction with the difference energy 143.756 MeV.
The energy of the $D_{sJ}$(2460)$^\pm$ $\rightarrow$ $D_s^\pm$ transition 
explains by the decay of the one complete 
$s$-oscillator with the energy 491.33 MeV. These results fully conform with
 experimental data of CLEO and BELLE (142.12~\cite{cleo}, 143.76~\cite{cleo2} 
and 491.4~\cite{belle} MeV).  
In present article the similar study of mass differences has been
continued for upper part of the charmonium spectrum, 
i.e. above $D\bar{D}$ threshold. 

The mass limiting is connected with feature of quark structures of $D$- and 
$c\bar{c}$-mesons~\cite{my2}. The lower part of charmonium spectrum perhaps
 is more difficult to an interpretation than upper part.  
The PDG data~\cite{parti} contains the next proper mesons: \\
$\psi$(3770), $X$(3872), $\chi_{c2}$(2P)(3929),  $X$(3940), 
$\psi$(4040), $\psi$(4160), $X$(4259) and $\psi$(4415).

Here we shall use the same labels as in~\cite{my3},  
the standard notation for the harmonic quarks ($d$, $u$, $s$, ...) 
and circle-enclosed characters for quarks of the completed harmonic oscillators
 (\cd, \cu, \cs, ...).    

The masses of harmonic quarks and completed harmonic 
oscillators are taken of~\cite{my3, my4}.

\section {The mass differences between $c\bar{c}$-mesons
and their interpretation}

\quad The experimental masses~\cite{parti} 
of above mentioned $c\bar{c}$-mesons are given in the table 1.

\par

{      
\begin{center}

Table 1. The experimental masses of some charmonium mesons.\\ 
\medskip\small

  \begin{tabular}{|c|c||c|c|}
   
 \hline
  Meson & Meson mass, &  Meson & Meson mass,\\
       &  MeV &  &  MeV \\
    \hline
 $\psi$(3770) & 3772.4 $\pm$1.1& $\psi$(4040) & 4039 $\pm$1 \\
 \hline
 $X$(3872) & 3871.4 $\pm$0.6& $\psi$(4160) & 4153 $\pm$3 \\
 \hline
 $\chi_{c2}$(2P)(3929) & 3929 $\pm$5&$X$(4259) & 4259 $\pm$3 \\
 \hline
 $X$(3940)& 3943 $\pm$17& $\psi$(4415) & 4421 $\pm$4 \\
\hline

  \end{tabular}

\end{center}
}

\par

All mass differences of mentioned mesons are given in table 2. 

\par

{      
\begin{center}

Table 2. The experimental mass differences (MeV) of charmonium mesons

 from $\psi$(3770) to $\psi$(4415).\\ 
  \medskip\small   \begin{tabular}{|c|c|c|c|c|c|c|c|}
    \hline
  Meson & $X$(3872) & $\chi$(3929) & $X$(3940) & $\psi$(4040) &
 $\psi$(4160)& $X$(4259) & $\psi$(4415) \\
  
\hline
 $\psi$(3770) & 99&157&171&{\bf 267}&{\em 381}&{\em 487}&649 \\
 \hline
 $X$(3872)    &////////&{\bf 58}&72&168&282&{\em 388}&{\bf 550} \\
 \hline
 $\chi$(3929) &&////////& 14&110&224&330&{\bf 492} \\
 \hline
 $X$(3940)    &&&////////& 96&{\bf 210}&{\bf 316}&478 \\
\hline
 $\psi$(4040) &&&&////////& 114&220&{\em 382} \\
 \hline 
 $\psi$(4160) &&&&&////////& {\bf 106}&{\bf 268} \\
 \hline
 $X$(4259)    &&&&&&////////& {\bf 162} \\
 \hline

  \end{tabular}

\end{center}
}

\par
\quad The author's data of simple quark groups are shown in the table 3.

{      
\begin{center}

Table 3. The rest energies of some quark pair and their completed oscillators.\\ 
  \medskip\small 
  \begin{tabular}{|c|c||c|c|}
    \hline
  Quark pair & Rest energy of quark pair & Quark pair  & Rest energy of quark pair\\
 or oscillator  & or oscillator, MeV     &or oscillator(s)  & or oscillator(s), MeV \\
    \hline
 $d\bar{d}$  & 57.62  & 2\cu\cub   & 268.50 \\
 \hline
  \cd\cdb    & 36.68  & 3\cu\cub   & 402.75 \\
 \hline
  $u\bar{u}$ & 210.88 & $s\bar{s}$ & 771.78 \\
 \hline
 \cu\cub     & 134.25 & \cs\csb    & 491.33 \\
\hline

  \end{tabular}

\end{center}
}

\par

Comparing data of table 2 with rest energies of quark pair and complete
oscillators (see table 3), we mark a lot of agreements. 
The full set of differences is equal 28.
Only part of them reflects simple physical reality, 
i.e. it is main differences which are directly connected with simple quark transitions. 
First they were discovered in work~\cite{my1}. This part was found large (\Approx half).
The number of obvious coincidences with masses of quarks 
or the elementary quark groups is equal to 13.
Exact coincident values in table 2 set off in {\bf bold} font, 
less exact - {\em italics}. 

The $\psi$(4415) has greatest number (4) of
exact differences. The $\psi$(3770) and $\psi$(4040) have only
one exact mass difference among themselves.
The analysis of data in table 2 has shown
that 5 of 7 mass differences with $\psi$(4415) 
are precisely interpreted.
Two remaining differences with $\psi$(3770) and $\psi$(4040) 
are not interpreted, so these mesons put into separate group.
The mass difference of these two mesons is equal to 267 MeV 
and it may be interpreted by two $u$-oscillators (268.5 MeV).
The mesons $\psi$(4415) and $\psi$(4160) have same mass difference, 
i.e. it is probably one of main differences.
Other main components were also chosen of data in table 2. 

Experimental mass differences of mesons and rest energies 
of interpreting quark groups are compared in table 4.
Experimental and calculated data are in the good consent, so good,
that it is possible even to note the overestimated error 
by which some experimental data are accompanied.

\par

{     
\begin{center}

Table 4. The calculated and experimental mass differences in upper part 
of the charmonium spectrum.\\ 
  \medskip\small 
  \begin{tabular}{|c|c|c|c|}
    \hline
  Mesons of difference &  Experimental mass& Quark presentation & Rest energy of \\
  & difference, MeV    &  & quark group, MeV \\
	   
    \hline
 $\psi$(4415) - $\chi$(3929) & 492 $\pm$6 & \cs\csb & 491.33  \\
 \hline
 $\psi$(4415) - $X$(4260) & 162 $\pm$10 & $du^0\bar{d}$ & 163.06 \\
 \hline
 $\psi$(4415 ) - $\psi$(4160) & 268 $\pm$5 & 2\cu\cub & 268.50 \\
 \hline
 $\psi$(4160) - $X$(3940) & 210 $\pm$8 & $u\bar{u}$ & 210.88 \\
 \hline
 
 $\chi$(3929) - $X$(3872) & 57.6 $\pm$5 & $d\bar{d}$ & 57.62 \\
 \hline

 $\psi$(4040) - $\psi$(3770) & 266.6 $\pm$1.5 & 2\cu\cub & 268.50 \\

 \hline

  \end{tabular}

\end{center}

}
\par

The interpretation of interlevel energies in upper part 
of the charmonium spectrum is shown also in figure 1. 

\par
\begin {figure} [htb]
\begin {center}
\includegraphics [scale=0.6] {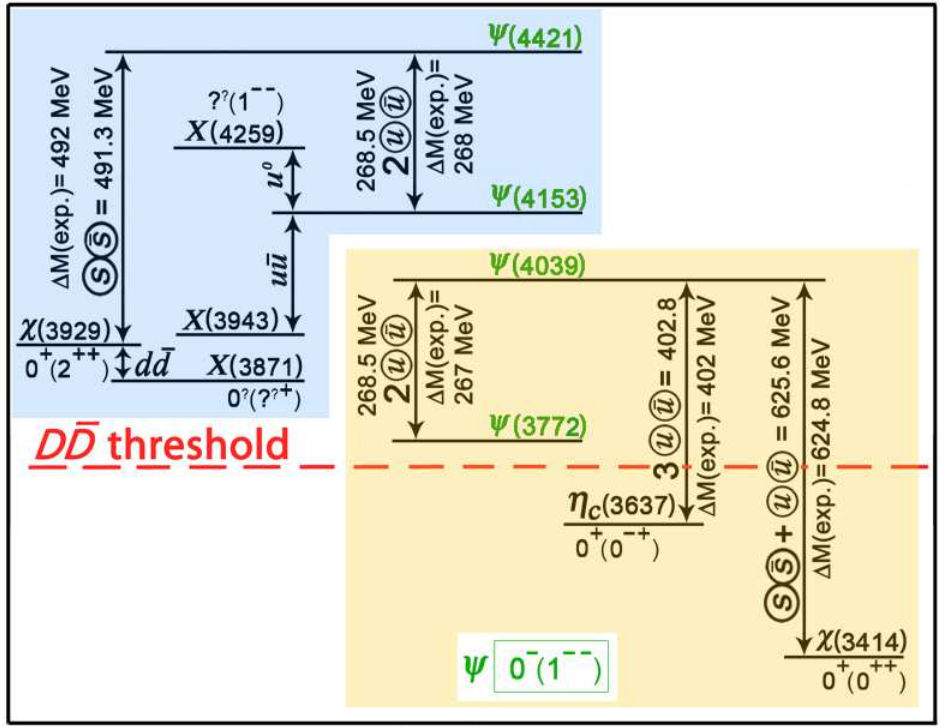}
\par
\caption [] {The interpretation  by harmonic quarks of some interlevel 
energies of the charmonium spectrum. 
Here in the brackets are given the experimental masses~\cite{parti}. 

\label {fig:charm}}
\end {center}
\end {figure}
\par

The spectrum consists of two groups. Except for two particles
connected among themselves, the others 5 particles 
are connected with a meson $\psi$(4415). 
The levels are separated by elementary groups of harmonic quarks
and oscillators. 

 All four $\psi$ have identical quantum numbers. 
They compose two separate pairs with equal interlevel energies 
and with exact interpretation by double $u$-oscillators.  
Among other levels in figure 1 there is just one transition 
with fully established changes of quantum numbers:
from $\psi$(4415) to $\chi$(3929). All odd quantum numbers 
change to even values. In addition spins of mesons change on 1. 
The interlevel energy is equal to one $s$-oscillator (491.33 MeV).
An aroma and a charge of particles do not change. 
Before in~\cite{my1} was discovered the transition 
from $D_s^\pm$(2460) to $D_s^\pm$(1968)  
with the same interlevel energy (491.4 MeV~\cite{belle}). 
Interestingly enough that it has the analogical change of quantum numbers. 
If $\psi$(4040) and $\psi$(3770) represents separate group 
we can expect similar transitions for top level ($\psi$(4040)).
 
In near part of charmonium spectrum there are only three possible transitions
with similar change of quantum numbers:

$\psi$(4040) $\rightarrow$ $\chi_{c0}$(3414);\quad \quad
$\psi$(4040) $\rightarrow$ $\chi_{c2}$(3556);\quad \quad
$\psi$(4040) $\rightarrow$ $\eta_c$(3637).
 
Here, the truth, we already get in area below $D\bar{D}$ a threshold. 
Corresponding mass differences are equal to 625, 483 and 402 MeV. 
Two of three energies can be interpreted immediately 
by two and three oscillators:

\cs\csb\, + \cu\cub = 625.6 (MeV) 
and 3\cu\cub = 402.75 (MeV).

Therefore harmonic quarks can good work and below threshold $D\bar{D}$.
These transitions of charmonium spectrum 
are shown also in figure 1.

\section {The recalculation of meson masses}  

\quad The quark interpretation of mass differences 
and most precisely measured masses of mesons $X$(3872) 
and $\chi_{c0}$(3414) were used for recalculation of other meson masses.
The mass error of $X$(3872) is equal to $\pm$0.6 MeV and it was used
for first meson gpoup ($\psi$(4415), $X$(4260), $\psi$(4160), 
$X$(3943) and $\chi$(3929)).
The $\chi_{c0}$(3414) mass has a smaller experimental error ($\pm$0.35 MeV) 
and so it was used as reference point for second group 
($\psi$(4040), $\psi$(3770) and $\eta_c$(3637). 
  
The calculated masses of mesons are given in table 5. 
The errors of calculated mass differences does not exceed 0.05 MeV 
and, hence, errors of the computed values correspond 
to accuracy of experimental data of mesons $X$(3872) and $\chi_{c0}$(3414),
i.e. properly $\pm$0.6 and $\pm$0.4 MeV.

{      
\begin{center}

Table 5. The calculated masses of some charmonium mesons.\\ 
  \medskip\small
  \begin{tabular}{|c|c|c|c|c|c||c|c|c|}
    \hline
  Mesons &$\psi$(4415) &$X$(4260)  &$\psi$(4160)&$X$(3940)&$\chi$(3929) & $\psi$(4040)  &$\psi$(3770)& $\eta_c$(3637) \\
 \hline
Calculated	&&&&&&&&\\
  mass, MeV &  4420.4   & 4257.3 &4151.8 &3941.0&3929.0& 4040.3&3771.8 &3637.6 \\
 
 \hline
Calculated  &&&&&&&& \\
 error, MeV & $\pm$0.6 & $\pm$0.6 & $\pm$0.6& $\pm$0.6  &  $\pm$0.6 & $\pm$0.4  & $\pm$0.4 & $\pm$0.4\\
    \hline

  \end{tabular}

\end{center}

}
\par

\section {Conclusion}

\quad It's discovered that the mass differences
of $c\bar{c}$-mesons are also rigorously quantized by the rest energies
of harmonic quarks and their completed oscillators.
The harmonic quarks and their derived structures are the building blocks
for the hadrons. They are perhaps the only building material.
Summarizing the results of the present work and works~\cite{my1, my2, my3}, 
it is possible to tell, that mesons with closed and opened charm are especially 
good objects for demonstration of real existence of harmonic quarks
and effectiveness of harmonic quark model.

\end {document}